\documentclass[%
aip,
reprint,
 amsmath,amssymb,
]{revtex4-1}

\usepackage{dcolumn}
\usepackage{bm}
\usepackage{graphicx}
\usepackage{hyperref}
\usepackage{siunitx}
\usepackage{natbib}
\usepackage[symbol]{footmisc}
\usepackage{footnote}
\usepackage{float}
\usepackage{color}
\setcitestyle{unsrtnat, square, comma, numbers}

\begin{document}

\title{Stabilizing zero-field skyrmions in Ir/Fe/Co/Pt thin film multilayers by magnetic history control}

\author{Nghiep Khoan Duong}
\affiliation{Division of Physics and Applied Physics, School of Physical and Mathematical Sciences, Nanyang Technological University, 21 Nanyang Link, Singapore 637371}

\author{M. Raju}
\affiliation{Division of Physics and Applied Physics, School of Physical and Mathematical Sciences, Nanyang Technological University, 21 Nanyang Link, Singapore 637371}

\author{A. P. Petrovi\'{c}}
\affiliation{Division of Physics and Applied Physics, School of Physical and Mathematical Sciences, Nanyang Technological University, 21 Nanyang Link, Singapore 637371}

\author{R. Tomasello}
\affiliation{Institute of Applied and Computational Mathematics, FORTH, GR-70013, Heraklion-Crete, Greece}

\author{G. Finocchio}
\affiliation{Department of Mathematical  and Computer Sciences,  Physical  Sciences  and Earth  Sciences, University of Messina, Messina 98166, Italy}

\author{Christos Panagopoulos}
\thanks{Corresponding author: christos@ntu.edu.sg}
\affiliation{Division of Physics and Applied Physics, School of Physical and Mathematical Sciences, Nanyang Technological University, 21 Nanyang Link, Singapore 637371}


\begin{abstract}
We present a study of the stability of room-temperature skyrmions in [Ir/Fe/Co/Pt] thin film multilayers, using the First Order Reversal Curve (FORC) technique and magnetic force microscopy (MFM). FORC diagrams reveal irreversible changes in magnetization upon field reversals, which can be correlated with the evolution of local magnetic textures probed by MFM. Using this approach, we have identified two different mechanisms - (1) skyrmion merger and (2) skyrmion nucleation followed by stripe propagation - \textcolor{black}{which facilitate magnetization reversal in a changing magnetic field.} Analysing the signatures of these mechanisms in the FORC diagram allows us to identify \textcolor{black}{magnetic ``histories'' - i.e. precursor field sweep protocols - capable of enhancing the final} zero-field skyrmion density. Our results indicate that FORC measurements can \textcolor{black}{play a useful role in} characterizing spin topology in thin film multilayers, \textcolor{black}{and are particularly suitable for identifying samples in which skyrmion populations can be stabilized at zero field.} 
\end{abstract}

\maketitle

Magnetic skyrmions are topologically charged spin textures which resist external perturbations and are thus appealing as potential information carriers for spintronic technology\cite{fert2013skyrmions,nagaosa2013topological, fert2017magnetic, soumyanarayanan2016emergent}. \textcolor{black}{Recently, it has been shown that \textcolor{black}{thin film multilayers comprising ferromagnets (FM) and heavy metals (HM)} are ideal candidates to host stable skyrmion phases\cite{bogdanov2001chiral}. In these materials, additively enhanced Dzyaloshinskii-Moriya interactions from neighbouring interfaces within the multilayers enable spin-winding (and hence, skyrmion formation) to occur even at room temperature (RT)\cite{woo2016observation, bode2007chiral, moreau2016additive, soumyanarayanan2017tunable}}. 

Successful implementation of magnetic skyrmions in technology calls for a thorough understanding of the magnetization reversal processes in the skyrmion-hosting materials. \textcolor{black}{The technique known as First Order Reversal Curve (FORC) acquisition provides an informative method of studying these processes.  FORC essentially} quantifies the degree of irreversible magnetization switching taking place as a magnetic material evolves under an applied field\cite{stancu2003micromagnetic, pascariu2012experimental, mayergoyz1988generalized, pike1999characterizing, roberts2000first, gilbert2014probing, roy2015exchange}. This approach has revealed various complex magnetization reversal behaviours, such as irreversible avalanche-like propagation of domains from previously nucleated bubbles in Co/Pt multilayers \cite{davies2004magnetization}, \textcolor{black}{and fractal domain propagation in CoCrPt thin films}\cite{navas2017microscopic}.

We anticipate that non-trivial magnetization reversals will also be exhibited in skyrmion-hosting materials, since skyrmion nucleation - like other processes involving modifications of domain topology - is magnetically irreversible\cite{bertotti1998hysteresis, olamit2007irreversibility, li2013nucleation}. Tracking the presence of irreversibility in these materials should hence enable history-dependent modifications to the remanent domain topology \cite{molho1987irreversible, westover2016enhancement}. Such behaviour could be utilised for stabilising zero-field skyrmions, which are technologically and scientifically attractive\cite{he2017realization, zheng2017direct, zhang2018direct}. This paper presents our study of magnetic irreversibility of skyrmions in FM/HM multilayers using the FORC technique. In particular, we examine the link between irreversible changes in local magnetization profiles at finite fields and the eventual presence of stable skyrmion populations at zero field.

Our choice of material in this study is the \big[Ir(10)/Fe(x)/Co(y)/Pt(10)\big]$_{20}$ thin film multilayer, where (10,x,y,10) are thicknesses in \si{\angstrom}. These heterostructures are known to exhibit magnetic skyrmions at RT (300 K) in the presence of an applied magnetic field\cite{soumyanarayanan2017tunable, raju2017evolution, yagil2018stray}. Varying the ferromagnetic (Fe(x)/Co(y)) composition allows magnetic interactions such as the Dzyaloshinskii-Moriya interaction (\textit{D}), exchange (\textit{A}) and magnetic anisotropy (\textit{K}) to be tuned. This in turn affects the thermodynamic stability of skyrmions, which is quantified by the parameter $\kappa$\cite{bogdanov1994thermodynamically,heide2008dzyaloshinskii,tomasello2018origin}:
\begin{equation}
\kappa = \frac{\pi}{4}\frac{D}{\sqrt{AK}}
\end{equation}
%

When $\kappa<1$, skyrmions exist as isolated, metastable entities. For $\kappa\geq1$, stable skyrmion lattices are expected to form in an out-of-plane magnetic field, although the equilibrium zero-field ground state remains a chiral spin-stripe phase\cite{bogdanov1994thermodynamically}. However, it is likely that the intrinsic skyrmion-skyrmion repulsion in dense arrays will allow individual topologically-charged skyrmions to resist deformation into stripes, and hence survive as the applied field is decreased. We therefore explore this possibility - stabilizing zero-field skyrmions using magnetic history control - in multilayers with compositions Fe(3)/Co(6) and Fe(4)/Co(6), whose $\kappa$ values - $1.1$ and $2.1$, respectively - have been shown to generate dense arrays of skyrmions in applied fields\cite{soumyanarayanan2017tunable}.

\begin{figure}
  \centering
  \includegraphics[width=9 cm,keepaspectratio]{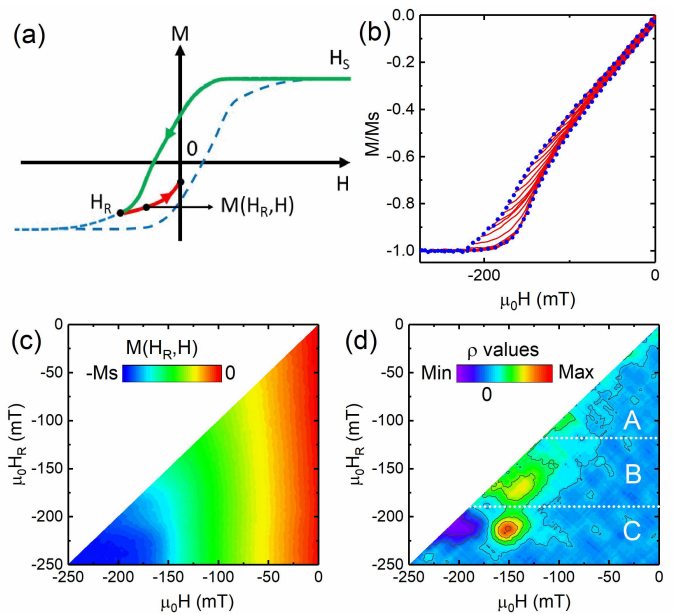}
  \caption{\textbf{FORC measurement procedures and magnetic irreversibilities in Fe(3)/Co(6)} (a) FORC measurement schematics: starting from a saturated state ($H_S$), the system is brought to a field $H_R$ (black dot at the end of the green line). A FORC (red line) is a reversal curve originating at $H_R$ on the major hysteresis loop, then rising to a final field (zero in our experiments). (b) A family of FORCs (red lines) is measured at equally spaced $H_R$, filling up the interior of the major hysteresis loop (blue dotted line). (c) 2D map of magnetization values traced out by the FORCs in (b) for different values of $H_R$ and $H$. (d) The resultant contour plot of $\rho$ values - known as the FORC diagram - calculated from the magnetization data shown in (c).}
  \label{fig:FORC-schematics}
\end{figure}

Our experiments use the same batch of samples, whose magnetic parameters (including $\kappa$) were previously characterized and discussed in ref. 9. We have employed similar magnetic force microscopy (MFM) techniques to image the spin textures. The FORC analysis we present here is based on a series of magnetization experiments, performed with a vibrating sample magnetometer (VSM) at RT. Each FORC measurement follows a two-part sequence. \textcolor{black}{First, a positive field ($400$ mT) sufficient to saturate the sample magnetization is applied perpendicular to the sample plane. This field is then swept down to a lower value, referred to as the reversal field, $H_R$ (\hyperref[fig:FORC-schematics]{Fig. 1(a)}). Next, the applied field is swept up from $H_R$ towards its final value (which is always $\mu_0H=0$ in our experiments), tracing a reversal branch known as a FORC} (red line, \hyperref[fig:FORC-schematics]{Fig. 1(a)}). The sequence is repeated for multiple values of $H_R$, equally spaced at $2.5$ mT-intervals between $-400$ and \textcolor{black}{$0$ mT}. This yields a family of FORCs which fill the interior of the major hysteresis loop(\hyperref[fig:FORC-schematics]{Fig. 1(b)}). The magnetization inside the major loop is described by a two-variable quasi-continuous function $M(H_R, H)$ (\hyperref[fig:FORC-schematics]{Fig. 1(c)}). From this, the degree of magnetic irreversibility can be quantified via the FORC distribution:
\begin{equation}
\rho = -\frac{1}{2}\frac{\partial ^2M(H_R, H)}{\partial H_R \partial H}
\end{equation} 

Purely reversible changes in magnetization give $\rho = 0$, while for irreversible changes $\rho \neq 0$\cite{roberts2000first}. A contour plot of the $\rho$ values \textcolor{black}{in the $(H,H_R)$ plane} (calculated by the FORCinel algorithms developed by \citet{harrison2008forcinel}) \textcolor{black}{is known as} a \textit{FORC diagram} (\hyperref[fig:FORC-schematics]{Fig. 1(d)}), which \textcolor{black}{locates and quantifies the} irreversibility of the magnetization reversals taking place in the sample\cite{roberts2000first, pike2003first}.  
A useful way to analyse a FORC diagram is to project $\rho$ onto the $H_R$ axis to obtain the switching field distribution (SFD). This is accomplished by integrating $\rho$ over H:
\begin{equation}
\rho_{SFD} = \int{-\frac{1}{2}\frac{\partial ^2M(H_R, H)}{\partial H_R \partial H}dH}
\end{equation} 

Plotting $\rho_{SFD}$ against $H_R$ \textcolor{black}{measures the irreversibility of the changes in magnetization which occur while sweeping the magnetic field back to zero from each reversal field $H_R$. In this way, distinct} irreversible processes taking place within different ranges of $H_R$ can be identified\cite{davies2005anisotropy, ruta2017first, winklhofer2006extracting}. The use of the $\rho_{SFD}$ plot will be demonstrated later in this paper. 
\begin{figure}
  \includegraphics[width=8.1cm, height = 8.1cm,keepaspectratio]{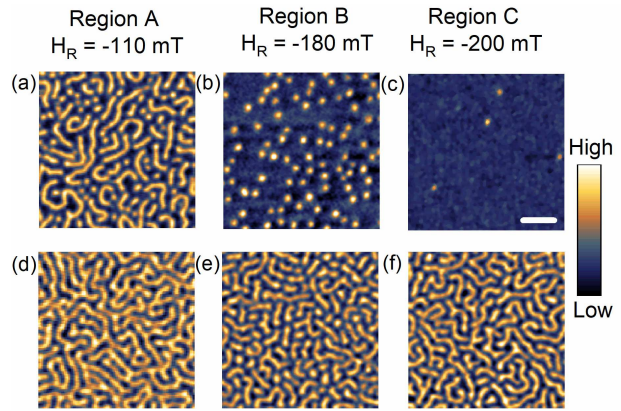}
  \caption{\textbf{Modification of domain topology in Fe(3)/Co(6)}. \textcolor{black}{(a)-(c) MFM images acquired at $\mu_0H=\mu_0H_R=$-110, -180 and -200~mT, respectively.  (d)-(f) MFM images subsequently measured at $\mu_0H=0$ following reversal from $H_R$} demonstrate clear modifications to the domain topologies. Fractured domains and isolated zero-field skyrmions are observed when $H_R$ falls into region B of the FORC diagram. Scale bar: 500 nm.}
  \label{fig:Fe3Co6-forc}
\end{figure}
\begin{figure*}
  \centering
  \includegraphics[width = 12.5cm,keepaspectratio]{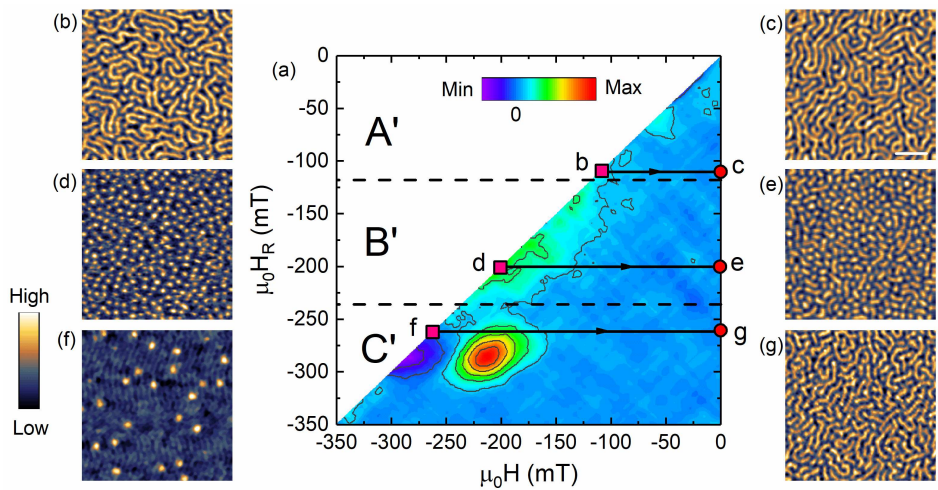}
  \caption{\textbf{Magnetic irreversibility and domain topology modification in Fe(4)/Co(6)}. (a) The FORC diagram of Fe(4)/Co(6) shows similar features in \textcolor{black}{$\rho(H,H_R)$ - labeled A$^{\prime}$, B$^{\prime}$ and C$^{\prime}$, respectively - to those of Fe(3)/Co(6). $H=H_R$ to $H=0$ reversal sequences are indicated by labeled arrows: b$\rightarrow$c (region A$^{\prime}$), d$\rightarrow$e (region B$^{\prime}$) and f$\rightarrow$g (region C$^{\prime}$).  (b)-(g) MFM images acquired before and after each reversal sequence.  A wide variety of domain topologies ranging from individual skyrmions to labyrinthine stripes can be stabilized in the remanent state (c,e,g), depending on the magnetic history.} Zero field skyrmions are notably observed at high density ($27\mu m^{-2}$) after sequence d$\rightarrow$e. Scale bar: 500 nm.}
  \label{fig:Fe4Co6-forc}
\end{figure*}

\textcolor{black}{We will focus on magnetization switching from negative values of $H_R$, where both skyrmions and stripes can be observed at RT. The} FORC diagram of Fe(3)/Co(6) reveals 3 major regions, labeled A, B and C as shown in \hyperref[fig:FORC-schematics]{Fig. 1(d)}. Region A, where $0\lesssim -\mu_0H_R\lesssim110$ mT, is a mostly featureless region with $\rho\sim0$. Magnetization changes in this region are therefore predominantly reversible. In contrast, a broad oval-shaped contour develops in region B ($110$ mT $\lesssim-\mu_0H_R\lesssim190$ mT), with positive $\rho$ values which peak at $\mu_0H_R\sim-175$ mT. This indicates that magnetic reversal in this $H_R$ range is an irreversible process, which may destroy pre-existing magnetic textures. Lastly, in region C, where $-\mu_0H_R\gtrsim190$ mT, which is close to the nominal saturation field of the sample, a pair of positive and negative $\rho$-contours emerge, both peaking at $\mu_0H_R\sim-210$ mT. 

Our MFM images show the presence of both stripes and skyrmions, especially at low magnetic fields. We distinguish these spin textures by their geometric eccentricity, taking account of our finite spatial resolution. Within a binarized image, the true eccentricity of a circular domain with radius $2\pm0.5$ pixels (the typical skyrmion size) can lie between 0 and 0.8.  Evaluating the eccentricity of all resolved spin textures in our images, we therefore set 0.8 as the threshold, below which the domain is considered a skyrmion, and otherwise a stripe. As the \textcolor{black}{reversal field amplitude rises} from region A to region C, MFM images obtained at $\mu_0H=\mu_0H_R$ indicate a gradual transition from a stripe phase to a skyrmion phase. Stripe domains, seen at $\mu_0H_R=-110$ mT (Region A, \hyperref[fig:Fe3Co6-forc]{Fig. 2(a)}), shrink to form circular skyrmions as the applied field increases to $\mu_0H_R=-180$ mT  (Region B, \hyperref[fig:Fe3Co6-forc]{Fig. 2(b)}). This is followed by a gradual annihilation of skyrmions \textcolor{black}{at higher fields (Region C, \hyperref[fig:Fe3Co6-forc]{Fig. 2(c)}). To examine what happens to these magnetic textures following magnetization reversals in each region, we reduce the applied field to zero from each $H_R$ and image the magnetic configurations of the remanent ($\mu_0H=0$) states}. In region A, reducing the magnetic field from $\mu_0H_R=-110$ mT to $\mu_0H=0$ mT converts the initially mixed stripe-skyrmion configuration into a labyrinthine network of stripe domains (\hyperref[fig:Fe3Co6-forc]{Fig. 2(d)}). In region B, however, the remanent state (reversed from $\mu_0H_R=-180$ mT) contains short, fractured stripe domains (\hyperref[fig:Fe3Co6-forc]{Fig. 2(e)}). Notably, some isolated circular skyrmions are seen to remain in the system. Lastly, for $\mu_0H_R=-200$ mT (\hyperref[fig:Fe3Co6-forc]{Fig. 2(f)}), the remanent state consists mostly of fractured stripe domains visibly longer than those in \hyperref[fig:Fe3Co6-forc]{Fig. 2(e)}. 

\textcolor{black}{The wide variety of domain configurations observed before and after magnetization reversal suggests that multiple reversal mechanisms may be active in skyrmion-hosting films.} Firstly, in region C of the FORC diagram, the positive-negative pair of $\rho$-contours is reminiscent of the high-field irreversibility signature ascribed to \textcolor{black}{bubble nucleation (and subsequent stripe propagation)} in Co/Pt multilayer thin films\cite{davies2004magnetization}. \textcolor{black}{In Fe(3)/Co(6), finite-length stripe domains are present at zero field after reversal from a state close to uniform ferromagnetic polarization. This suggests that the primary magnetization reversal mechanism in region C is the (irreversible) nucleation of skyrmions and their ensuing (reversible) propagation into stripes}. Secondly, the presence of a dense array of skyrmions at $\mu_0H=\mu_0H_R$ coincides with the $\rho$-contour arising in region B, \textcolor{black}{where a mixed population of skyrmions and short stripes survives reversal to zero field. Here, the likely source of irreversible magnetization switching is} the merger of skyrmions to form stripe domains: as the magnetic field is reduced, skyrmions become unstable to elliptical deformations and expand until they collide and merge with a neighbouring skyrmion or stripe \cite{bogdanov1994thermodynamically}. Such mergers are irreversible, for they require skyrmions to unwind their topologically protected spin textures to form a single stripe - hence, topological charge is not conserved. Lastly, when the system is in region A of the FORC diagram, \textcolor{black}{many stripes remain present at $\mu_0H=\mu_0H_R$ and the skyrmion density is low.  Therefore, less topological charge annihilation is required for magnetization reversal, accounting} for the smaller values of $\rho$ in this region. 

Our observations indicate that two main reversal mechanisms take place in Fe(3)/Co(6): (1) \textcolor{black}{skyrmion nucleation at high fields near saturation (region C) followed by reversible propagation into stripes, and (2) skyrmion merger at lower reversal fields within the skyrmion phase (region B). In region B, although many skyrmions appear to have merged into stripes at zero field after reversal, some skyrmions remain visible (\hyperref[fig:Fe3Co6-forc]{Fig. 2(e)}). While this can be partially attributed to disorder-induced pinning in our sputtered films, the survival of skyrmions at zero field is also linked to their thermodynamic stability in Fe(3)/Co(6), where $\kappa=1.1$\cite{soumyanarayanan2017tunable}. It is plausible that an even higher density of zero-field skyrmions could be achieved in samples with larger $\kappa$ by following similar field-reversal protocols. We therefore conduct the same set of experiments - FORC analysis and MFM imaging - on sample Fe(4)/Co(6) ($\kappa=2.1$), in order to test this hypothesis and validate our proposed magnetization reversal mechanisms}.
\begin{figure}
  \includegraphics[width=8.5cm, height =9cm,keepaspectratio]{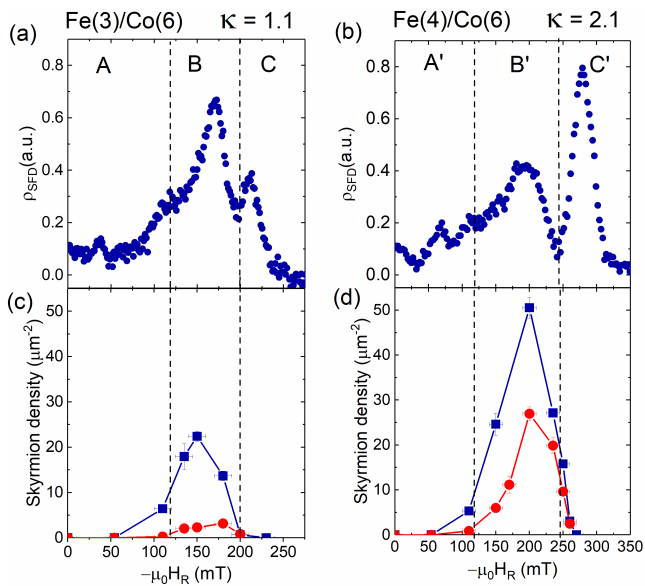}
  \caption{\textbf{Normalized switching field distributions (SFD) and zero-field skyrmion stability for Fe(3)/Co(6) and Fe(4)/Co(6)}. (a,b) Normalized SFDs for both samples display two peaks: one associated with skyrmion merger (B,B$^{\prime}$), the other corresponding to skyrmion nucleation (C,C$^{\prime}$) following magnetization reversals. (c,d) Skyrmion densities measured by MFM at $H=H_R$ (blue lines) and $H=0$ (red lines) after reversing from $H_R$. In Fe(4)/Co(6), the maximum skyrmion densities achievable at $H_R$ and zero-field are $50$ and $27$  $\mu$m$^{-2}$ respectively, compared with $23$ and $3$  $\mu$m$^{-2}$ in Fe(3)/Co(6).} 
  \label{fig:SFD-SkN}
\end{figure}

The FORC diagram for Fe(4)/Co(6) is shown in \hyperref[fig:Fe4Co6-forc]{Fig. 3(a)}: magnetization reversals in this sample exhibit features consistent with those of Fe(3)/Co(6). The key regions are labelled A$^{\prime}$, B$^{\prime}$ and C$^{\prime}$, in correspondence with the A, B and C regions in \hyperref[fig:FORC-schematics]{Fig. 1(d)}. \textcolor{black}{Similar to Fe(3)/Co(6), the magnetic configuration at $\mu_0H=\mu_0H_R$ evolves from a mixed stripe-skyrmion phase (A$^{\prime}$, \hyperref[fig:Fe4Co6-forc]{Fig. 3(b)}) to a full skyrmion phase (B$^{\prime}$, \hyperref[fig:Fe4Co6-forc]{Fig. 3(d)}) and finally a near-saturation phase where most skyrmions have already been annihilated (C$^{\prime}$, \hyperref[fig:Fe4Co6-forc]{Fig. 3(f)}). Following magnetization reversal, short fractured domains (including skyrmions)} are observed at remanence (\hyperref[fig:Fe4Co6-forc]{Fig. 3(c),(e),(g)}). Remarkably, the skyrmion density at zero field for Fe(4)/Co(6) reaches a maximum of $\sim27$ ${\mu}${m}$^{-2}$, compared to $50$ $\mu$m$^{-2}$ attained at the reversal field \hyperref[fig:SFD-SkN]{Fig. 4(d)}. In contrast, zero-field skyrmion density can only reach $\sim3$ ${\mu}${m}$^{-2}$ in Fe(3)/Co(6). 

\textcolor{black}{The emergence of such a large zero-field skyrmion population in high-$\kappa$ multilayers can be understood by evaluating $\rho_{SFD}$ in both of our samples (\hyperref[fig:SFD-SkN]{Fig. 4(a),(b)}).  Peaks indicating irreversible changes in magnetization emerge in regions B,B$^{\prime}$ and C,C$^{\prime}$ for both films; however, the separation and relative amplitude of these peaks differ strongly between samples.  For Fe(3)/Co(6), the peaks in B and C are separated by $\approx40$~mT and the peak in B is dominant.  In contrast, for Fe(4)/Co(6), the peaks in B$^{\prime}$ and C$^{\prime}$ are separated by $\approx90$~mT and the peak in C$^{\prime}$ is far larger.  Recalling the principal reversal mechanisms occurring within each region of the FORC diagrams, we deduce that the peaks in C,C$^{\prime}$ correspond to skyrmion nucleation, while those in B,B$^{\prime}$ are caused by topological charge annihilation via skyrmion merger.} We note that skyrmion depinning (and subsequent motion as the field is reduced) may also contribute to the magnetic irreversibility observed in regions (B,B$^{\prime}$). Such behaviour is analogous to the well-known irreversibility caused by domain wall depinning \cite{jiles1986theory}. However, the correlation between the decrease in skyrmion density upon field reversal and the rise in peak amplitude of $\rho_{SFD}$ within (B,B$^{\prime}$) indicates that skyrmion merger is the dominant contributor to irreversibility in this field regime. 

\textcolor{black}{The qualitative trends linking $\rho_{SFD}(H_R)$ with our observed zero-field magnetization textures are clear. In Fe(3)/Co(6), skyrmions are less likely to form (lower peak in C) and more likely to merge into stripes (higher peak in B).  Conversely, in Fe(4)/Co(6) skyrmions can be} \textcolor{black}{stabilised} \textcolor{black}{within a broader range of magnetic fields due to their higher $\kappa$, creating a larger separation between the peaks in B$^{\prime}$ and C$^{\prime}$. Skyrmions are therefore more likely to form (higher peak in C$^{\prime}$) and less likely to merge into stripes (lower peak in B$^{\prime}$).} Furthermore, due to the higher $\kappa$ value, the skyrmion radius decreases and a denser array of skyrmions is formed at the reversal field $H_R$. These small, closely-packed skyrmions experience comparatively stronger repulsive forces from their neighbours\cite{lin2013particle, zhang2015skyrmion}, preventing them from merging into stripes. Consequently, the density of skyrmions which can be engineered to survive at zero field is far higher in Fe(4)/Co(6), in agreement with our MFM data (\hyperref[fig:SFD-SkN]{Fig. 4(c),(d)}).  

It is difficult to draw more quantitative conclusions from our $\rho_{SFD}$ plots, since the absolute peak magnitudes may be affected by mean-field and magnetostatic interactions\cite{ruta2017first, muxworthy2004influence}, especially in regions C,C$^{\prime}$ where negative FORC contours are observed.  Nevertheless, our results show that FORC diagrams and associated $\rho_{SFD}(H_R)$ curves can predict the propensity of ferromagnetic multilayers to host stable zero-field skyrmions, even without any knowledge of $\kappa$ or prior imaging experiments.  Moreover, our data indicate a simple route to maximize zero-field skyrmion densities: field reversal from an intermediate (i.e. region B,B$^{\prime}$) value of $H_R$.

In summary, our FORC \textcolor{black}{and MFM} measurements reveal two distinct irreversible processes contributing to the magnetization reversals in the [Ir/Fe/Co/Pt]$_{20}$ multilayers: (1) skyrmion nucleation from a uniformly polarized state and (2) skyrmion mergers forming stripe domains. These processes are controllable by magnetic field cycling which modifies the remanent domain topology, resulting in the presence of fractured domains and isolated skyrmions at zero field. \textcolor{black}{We have shown how FORC diagrams and associated SFD plots can be used to establish field cycling protocols for maximizing zero-field skyrmion densities, as well as for inferring trends in skyrmion stability between different multilayer compositions.  FORC analysis should therefore be considered a valuable tool for characterizing and tuning spin topology in non-collinear magnets.} 

We thank Anthony Tan and James Lourembam for experimental inputs in the early stages of this work. This work was supported by the CN Yang Scholars Programme, the Singapore Ministry of Education (MoE), Academic Research Fund Tier 2 (Ref. No. MOE2014-T2-1-050) and the National Research Foundation (NRF) of Singapore, NRF – Investigatorship (Ref. No. NRF-NRFI2015-04). R.T. and G.F. thank the project “ThunderSKY” funded from the Hellenic Foundation for Research and Innovation and the General Secretariat for Research and Technology, under Grant No. 871.

\end{document}